\documentclass[aps,pra,twocolumn,showpacs,superscriptaddress]{revtex4-2}

\usepackage{graphicx}
\usepackage{amssymb}
\usepackage{amsmath}
\usepackage{color,xcolor}
\usepackage{bm}
\usepackage{physics}
\usepackage{ulem}

\usepackage{changes}

\newcommand{\quoting}[1]{``#1''}

\newcommand{\chk}[1]{\textcolor{black}{#1}}

\begin{document}


\title{A unified framework for exceptional point pairs in non-Hermitian two-level systems}   

\author{Jung-Wan Ryu}
    \address{Center for Theoretical Physics of Complex Systems, Institute for Basic Science (IBS), Daejeon 34126, Republic of Korea}
    \address{Basic Science Program, Korea University of Science and Technology (UST), Daejeon 34113, Republic of Korea}
\author{Jae-Ho Han}
\email{\chk{jaehohan@kaist.ac.kr}}
    \address{\chk{Department of Physics, Korea Advanced Institute of Science and Technology (KAIST), Daejeon 34141, Republic of Korea}}
\author{Chang-Hwan Yi}
\email{yichanghwan@hanmail.net}
    \address{Center for Theoretical Physics of Complex Systems, Institute for Basic Science (IBS), Daejeon 34126, Republic of Korea}
\date{\today}

\begin{abstract}
Exceptional points (EPs) in non-Hermitian systems are branch singularities where eigenvalues and eigenvectors simultaneously coalesce, leading to rich topological phenomena beyond those in Hermitian systems. In this work, we systematically investigate the interplay between eigenenergy braiding and Berry phase accumulation in two-level non-Hermitian systems hosting pairs of EPs. EP pairs are classified into four distinct classes according to the vorticity of eigenenergies, the Berry phase accumulated during encircling, and the eigenstate projection onto a basis state. Their associated topological structures are analyzed using effective two-level models. These classifications are further substantiated by numerical simulations in optical microcavities with three scatterers, where EPs emerge in the complex frequency spectrum. By encircling different EP pairs in parameter space, we demonstrate that the resulting topological features such as trivial or non-trivial braiding and Berry phase accumulation are directly linked to the vorticity structure and eigenmode evolution. In particular, we show that the eigenstate projection onto a basis state near EPs manifests as chiral optical modes in microcavities, providing an experimentally accessible signature of the underlying topological structure. Our results provide a unified framework for understanding multi-EP topology and offer practical pathways toward their realization and control in photonic systems.
\end{abstract}

\maketitle

\section{Introduction}

Non-Hermitian systems have emerged as a fertile ground for exploring novel topological phenomena that are absent in Hermitian physics \cite{Moiseyev_2011, El-Ganainy2018non, Ashida2020non, Bergholtz2021exceptional}. Among their most distinctive features are exceptional points (EPs), at which eigenvalues and eigenvectors simultaneously coalesce, thereby forming non-Hermitian branch points \cite{Kato1976perturbation, Heiss1990avoided, Ryu2009coupled, Dietz2011exceptional, Xu2016topological, Miao2016orbital, Ding2016emergence, Chen2017exceptional, Oezdemir2019parity, Miri2019exceptional}. The adiabatic encirclement of EPs gives rise to unique topological effects, including nontrivial eigenenergy braiding and Berry phase accumulation, which have been extensively studied both theoretically and experimentally \cite{Heiss1999phases, Dembowski2001experimental, Dembowski2004encircling, Gao2015Observation, Ryu2012analysis, Ryu2012geometric, Zhong2018winding}.

In two-level non-Hermitian models, encircling a single EP typically results in a square-root branch cut structure for the eigenenergies, inducing a permutation of eigenstates along with the accumulation of a $\pi$ Berry phase. These two intertwined phenomena, eigenenergy braiding and Berry phase accumulation, highlight that the topological characterization of EPs cannot rely solely on eigenvalue behavior. Instead, the internal structure of eigenstates and their associated geometric phases must also be incorporated to fully capture the underlying topology.

Multiple EPs can arise in general non-Hermitian systems, leading to rich topological structures and phenomena \cite{Luitz2019exceptional, Zhong2023numerical, Erb2024novel}.
When two EPs among many are jointly encircled along a closed loop, the eigenenergies may exhibit either a trivial or non-trivial braid \cite{Ryu2025complex}, and the Berry phase accumulated along the path can be either trivial or non-trivial ($0$ or $\pi$), depending on the relative chiralities of the eigenenergies and eigenstates at the EPs \cite{Heiss2012the, Zhou2018observation, Su2021direct, Ferrier2022unveiling, Zhang2022universal, Ryu2024exceptional, Kozii2024non, Ryu2024realization}.
This subtle interplay reveals that eigenenergy braiding and Berry phase accumulation are distinct yet fundamentally connected topological features of non-Hermitian topology.

In this work, we focus on two-level non-Hermitian systems containing two EPs and systematically explore how eigenenergy braiding and Berry phase accumulation combine when the parameter space is traversed adiabatically along closed loops. We aim to provide a comprehensive framework to understand the topological interplay between eigenvalue structure and geometric phases, and to suggest pathways for extending these ideas toward multi-EP systems.

\section{Topological structures of complex eigenenergies and eigenstates associated with EPs}
\label{sec:theory}

In this work, we classify EP pairs into four distinct classes. The classification is based on three key criteria: (i) the vorticities and braids of the eigenenergies, (ii) the Berry phase accumulated during encircling, and (iii) the projection of eigenstates onto basis states near the EPs. Accordingly, EP pairs with opposite vorticities are divided into Class I and Class II, while EP pairs with the same vorticities are divided into Class III and Class IV, depending on their Berry phase and eigenstate projection characteristics.

\subsection{A $2 \times 2$ Hamiltonian with an EP and topological invariants}
\label{sec:oneEP}

When a Hamiltonian hosts an EP, its local behavior near the EP can be captured by a Jordan normal form parametrized by a complex variable \( z \) representing deviations from the EP. In such a case, the Hamiltonian can be locally expressed as
\begin{align}
\label{eq:h0}
H(z) = \begin{pmatrix}
0 & 1 \\
z - z_1 & 0
\end{pmatrix},
\end{align}
where an EP locates at \( z=z_1 \). The non-diagonalizability at the EP characterizes the coalescence of both eigenenergies and eigenstates. This local description provides a minimal and universal structure to analyze the singular behavior of non-Hermitian systems near EPs, capturing both the algebraic and topological features intrinsic to EPs.

Two complex eigenenergies and their corresponding eigenstates coalesce at an EP, around which the eigenenergies $E_{\pm}$ exhibit a square root branch behavior, 
\begin{equation}
E_{\pm} = \pm \sqrt{z - z_1} ,
\label{eq:oneEP}
\end{equation}
where the complex parameter $z$ represents two-dimensional coordinates [Re($z$), Im($z$)] and $z_1$ is the parameter set of the EP in non-Hermitian systems. If we consider a closed loop $\Gamma$ encircling the EP in the parameter space $\bm{\alpha}$, the vorticity $\nu_1(\Gamma)$ can be defined as follows \cite{Leykam2017edge, Shen2018topological}: 
\begin{equation}
\nu_{1} (\Gamma) = - \frac{1}{2 \pi} \oint_{\Gamma} \nabla_{\bm{\alpha}} \mathrm{arg}[E_{+} (\bm{\alpha}) - E_{-} (\bm{\alpha})] \cdot d \bm{\alpha} . 
\label{eq:vorticity}
\end{equation}
Here, $\nu_1$ corresponds to the winding number of the eigenenergies $E_{+}$ and $E_{-}$ around the EP in the complex-energy plane. By reparametrizing the coordinate as $z - z_1 = r e^{i \theta}$, such that $E_{\pm} = \pm \sqrt{r} e^{i \theta / 2}$, one finds that the winding number is given by $\nu_1 = \pm 1/2$ for a closed loop $\Gamma$ encircling an EP. The \quoting{$\pm$} signs of $\nu_1$ indicate counterclockwise and clockwise rotations on the complex-energy plane, respectively, depending on the sign of $\theta$. 
In two-state models, the vorticity is equivalent to the winding number of the braid, defined as \cite{Gong2018topological}:
\begin{equation}
    w = \sum_{n=1}^{2} \oint_{\mathcal C(\lambda)} \frac{d\lambda}{2\pi} \frac{d}{d\lambda} \arg E_n(\lambda),
\end{equation}
which captures how the eigenenergies evolve as the system parameter traces a closed loop around a reference point, such as an EP considered in this work. $\mathcal C(\lambda)$ is a closed path in parameter space of the Hamiltonian, parameterized by $\lambda \in [0,1]$ with $\mathcal C(0) = \mathcal C(1)$. While encircling an EP, the winding number (braid degree) $w$ equals to $\pm 1$ and corresponding vorticities of each states are $\pm 1/2$. We note that the vorticity characterizes the local fractional winding of each eigenenergy around a single EP, whereas the braid degree denotes the global integer winding of all eigenenergies that reflects their braid structure. 

Meanwhile, the two eigenstates of Eq.~(\ref{eq:h0}) corresponding to the eigenvalues \( E_\pm \) are
\begin{equation}
\ket{\psi} =
\begin{pmatrix}
\pm \sqrt{(z-z_1)} \\
1
\end{pmatrix}.
\label{oneEPHamiltonian}
\end{equation}
By computing the phase evolution of these eigenstates under adiabatic variation of the parameter along a closed loop, we can examine whether a Berry phase is accumulated. The complex Berry phase in non-Hermitian Hamiltonians
is defined as \cite{Garrison1988complex, Dattoli1990geometrical, Mostafazadeh1999a, Liang2013topological}:
\begin{equation}
    \gamma = i \oint_{\mathcal C(\lambda)} \frac{\left< \phi (\lambda) | \partial_\lambda \psi (\lambda)\right>}{\left< \phi (\lambda) | \psi (\lambda)\right>} d\lambda.
\label{Berryphase}
\end{equation}
Here, $\left|\phi\right>$ and $\left|\psi\right>$ are left and right eigenstate of the Hamiltonian, respectively. The accumulated phase is $\pi$ for non-separated states associated with an EP. At the EP, one component of the eigenstate dominates. That is, the eigenstate is $(1,0)^T$ when $z = z_1$. We refer to this as the projection of the eigenstate onto a basis state.

\subsection{A $2 \times 2$ Hamiltonian with two EPs}
\label{sec:twoEPs}

In the following, we consider a closed loop that encircles two EPs. There are two types of EP pairs \cite{Ryu2025complex}: counter-rotating EPs, which possess opposite vorticities, and co-rotating EPs, which share the same vorticity. These are referred to as type-I and type-II EP pairs, respectively. The corresponding eigenenergy structures are topologically equivalent to
\begin{align}
\label{eq:type1}
    E_{\pm}^{\mathrm{I}} &= \pm \sqrt{( z - z_1 )( z - z_2 )^*}, \\
\label{eq:type2}
    E_{\pm}^{\mathrm{II}} &= \pm \sqrt{( z - z_1 )( z - z_2 )},
\end{align}
where $ z_1 $ and $ z_2 $ denote the positions of the two EPs, and the asterisk symbol (*) represents the complex conjugate.

\subsubsection{Type-I EP pairs: Class I}
Assuming that the two EPs in a type-I pair are sufficiently close compared to the radius of the encircling loop, one can consider two types of effective Hamiltonians that reproduce the energy spectrum given in Eq.~(\ref{eq:type1}). The first is given by
\begin{equation}
 \mathbf{Class~I:}~   H(z) = 
\begin{pmatrix} 
 0 & z-z_1 \\
 (z-z_2)^* & 0 \\
\end{pmatrix} ,
\label{eq:model_1}
\end{equation}
where $z = x + i y$ is a two-dimensional parameter, and there are two EPs at 
$z=z_1$ and $z=z_2$. 
If we set $z = r e^{i \theta}$ with
$r \gg |z_1|,\, |z_2|$, the eigenenergies and the corresponding left and right eigenstates can be approximated as
\begin{equation}
    E_{\pm}=E^\text{I}_\pm \approx \sqrt{z z^*} = \pm r ,
\label{eq:model_1b_ev}
\end{equation}
\begin{equation}
   \bra{\phi(\theta)} \approx 
\begin{pmatrix} 
 \pm e^{-i \theta}& 1 \\
 \end{pmatrix},\
    \ket{\psi(\theta)} \approx 
\begin{pmatrix} 
 \pm e^{i \theta} \\
 1 \\
\end{pmatrix}.
\label{eq:model_1b_es}
\end{equation}
It implies that the braid degrees are $0$ (i.e., no $\theta$-dependence of $E_\pm$) and accumulated Berry phases are $\pi$ during $\theta$ changes from zero to $2 \pi$ [see Eq.~(\ref{Berryphase})] with fixed $r$. In this case, if $z_1$ coincides with $z_2$, the two EPs merge into a diabolic point. Meanwhile, as $z$ approaches either $z_1$ or $z_2$, the eigenstates project onto a basis state, $(0,1)^T$ or $(1,0)^T$, respectively. The projections onto a basis state can manifest, for example, as chiral modes rotating in clockwise (CW) and counter-clockwise (CCW) directions in asymmetric microcavities.

\subsubsection{Type-I EP pairs: Class II}
Next, we consider the Hamiltonian with the same eigenenergies but is different from the class I.
\begin{equation}
 \bold{Class~II:}~   H(z) = 
\begin{pmatrix} 
 0 & 1 \\
 (z-z_1)(z-z_2)^* & 0 \\
\end{pmatrix} .
\label{eq:model_2}
\end{equation}
Under the same assumption, $r \gg |z_1|,\, |z_2|$ for $z = r e^{i \theta}$, we get

\begin{equation}
    E_{\pm} =E^\text{I}_\pm\approx \pm r ,
\label{eq:model_1b_ev}
\end{equation}
\begin{equation}
   \bra{\phi} \approx 
\begin{pmatrix} 
 \pm r & 1 \\
 \end{pmatrix},\
    \ket{\psi} \approx 
\begin{pmatrix} 
 \pm r \\
 1 \\
\end{pmatrix} .
\label{eq:model_1b_es}
\end{equation}
As in class I, the braid degrees are 0. However, in this case, the accumulated Berry phases also vanish for a fixed $r$. Furthermore, if $z_1$ coincides with $z_2$, the eigenstates project onto a basis state $(1,0)^T$ as $z$ approaches $z_1$ (or equivalently $z_2$).

\subsubsection{Type-II EP pairs: Class III}
In the case of type-II EP pairs, we also consider two different simplified effective Hamiltonians which have the same eigenenergies, Eq.~(\ref{eq:type2}),
\begin{equation}
 \bold{Class~III:}~   H(z) = 
\begin{pmatrix} 
 0 & z-z_1 \\
 z-z_2 & 0 \\
\end{pmatrix} .
\label{eq:model_3}
\end{equation}
For $z = r e^{i \theta}$ and $r \gg |z_1|,\, |z_2|$, the eigenenergies and the eigenstates are given as follows: 
\begin{equation}
    E_{\pm}(\theta) = E_{\pm}^{\mathrm{II}}\approx \sqrt{z^2} = \pm re^{i \theta} , 
\label{eq:model_1b_ev}
\end{equation}
\begin{equation}
    \bra{\phi} \approx 
\begin{pmatrix} 
 \pm 1 & 1
\end{pmatrix}, \
    \ket{\psi} \approx 
\begin{pmatrix} 
 \pm 1 \\
 1 \\
\end{pmatrix} .
\label{eq:model_1b_es}
\end{equation}
Unlike type-I, the braid degrees are now $2$ (i.e., $E_\pm$ is dependent on $\theta$), and accumulated Berry phases are $0$ during $\theta$ changes from zero to $2 \pi$ with fixed $r$. In this case, if $z_1$ and $z_2$ are the same, two EPs merge into a vortex point \cite{Shen2018topological}, and the eigenstates project onto a basis state, $(0,1)^T$ or $(1,0)^T$ as $z$ approaches $z_1$ or $z_2$. 

\subsubsection{Type-II EP pairs: Class IV}
In the following, we examine the final class in which the Hamiltonian is distinct from that of Class III, while yielding identical eigenenergies.
\begin{equation}
 \bold{Class~IV:}~   H(z) = 
\begin{pmatrix} 
 0 & 1 \\
 (z-z_1)(z-z_2) & 0 \\
\end{pmatrix} .
\label{eq:model_4}
\end{equation}
The eigenenergies and the eigenstates are now
\begin{equation}
    E_{\pm}(\theta) = E_{\pm}^{\mathrm{II}}\approx \pm re^{i \theta},
\label{eq:model_4b_ev}
\end{equation}
\begin{equation}
\bra{\phi(\theta)} \approx 
\begin{pmatrix} 
 \pm re^{i \theta} & 1
\end{pmatrix}, \
    \ket{\psi(\theta)} \approx 
\begin{pmatrix} 
 \pm re^{i \theta} \\
 1 \\
\end{pmatrix},
\label{eq:model_4b_es}
\end{equation}
under the assumption that $r \gg |z_1|,\, |z_2|$ for $z=r e^{i \theta}$. Therefore, the braid degrees are $2$ and accumulated Berry phases are $\pi$ during $\theta$ changes from zero to $2 \pi$ with fixed $r$. For $z_1=z_2$, the eigenstates project onto a basis state $(1,0)^T$ as $z$ approaches $z_1$ or $z_2$. The four classes are summarized in Table~\ref{tab01}.

\begin{table}
\centering
    \begin{tabular}{l|cccc}
    \hline
        EP pairs & Class I & Class II & Class III & Class IV \\
    \hline
        (i) Vorticities & ($\pm$,$\mp$) & ($\pm$,$\mp$) & ($\pm$,$\pm$) & ($\pm$,$\pm$) \\
        (ii) Braid degrees & $0$ & $0$ & $\pm 2$ & $\pm 2$ \\
        (iii) Berry phases & $(\pi,\pi)$ & (0,0) & (0,0) & $(\pi,\pi)$ \\ 
        (iv) Projection onto & & & & \\
        ~~~~~~a basis state & Opposite & Same & Opposite & Same \\
        ~~~~~~at two EPs & & & & \\
        (v) Merging & DP & EP & VP & EP \\
    \hline
    \end{tabular}
\caption{\label{tab01} Classes of EP pairs for two-level Hamiltonians. Each class is defined by three criteria: (i, ii) the vorticities of the two EPs and corresponding braids, (iii) the Berry phases for two states accumulated during encircling, and (iv) the eigenstate projection onto basis states at the EPs. Accordingly, Type-I EP pairs (opposite vorticities) correspond to Class I and Class II, while Type-II EP pairs (same vorticities) correspond to Class III and Class IV. Column (v) indicates the phase when two EPs merge: diabolic point (DP), vortex point (VP), or defective EP.}
\end{table}

\subsection{Generalization for Multiple EPs}
\label{sec:multiEPs}
The extension from two to multiple EPs follows straightforwardly in two-level systems, as the key mechanism for forming topological structures associated with EPs remains consistent. Let us consider a system with $N$ EPs, whose eigenenergies are given by
\begin{equation}
\label{eq:nEPs}
E_{\pm} = \pm \sqrt{( z - z_1 )^{(*)}( z - z_2 )^{(*)}( z - z_3 )^{(*)} \cdots ( z - z_N )^{(*)}}. \nonumber
\end{equation}
Here, the asterisk in the parenthesis \quoting{$(*)$} indicates the possibility of complex conjugation. The corresponding effective Hamiltonian can generally be written as
\begin{gather}
H(z) = \nonumber \\
\begin{pmatrix} 
 0 & (z-z_{i1})^{(*)}(z-z_{i2})^{(*)} \cdots \\
 (z-z_{j1})^{(*)}(z-z_{j2})^{(*)} \cdots & 0
\end{pmatrix}.
\label{eq:model_multiEPs}
\end{gather}
Each EP located at $z = z_{in}$ or $z = z_{jn}$ leads to eigenstates that project onto a basis state, either $(0,1)^T$ or $(1,0)^T$, respectively. The vorticities, braid degrees, and accumulated Berry phases associated with encircling multiple EPs can be obtained using the same methods as in the two-EP case.

\begin{figure*}
    \centering
    \includegraphics[width=1.0\linewidth]{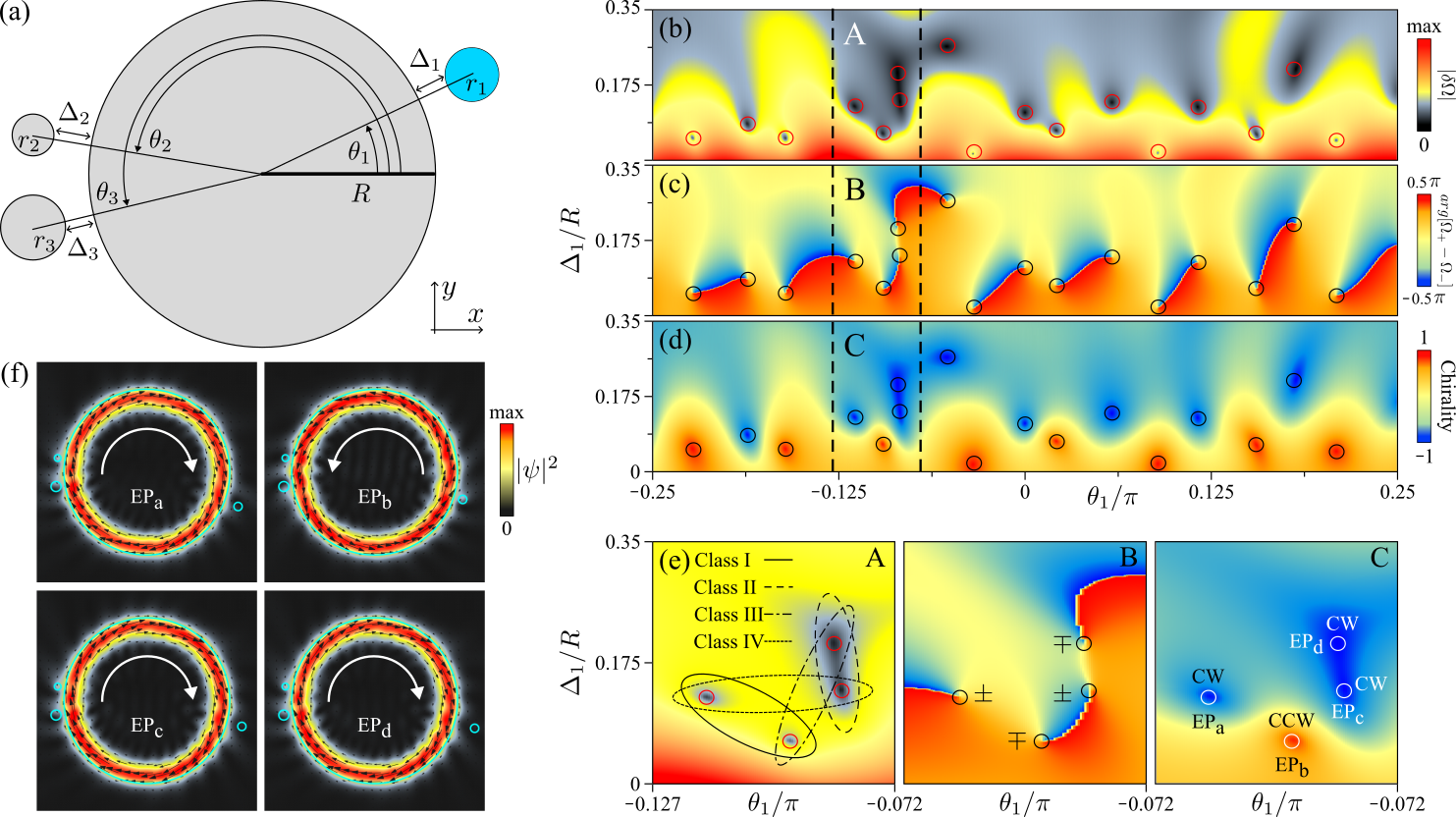}
    \caption{(a) Schematic of a dielectric microdisk with three scatterers placed outside its boundary. The values of $r_i$ and $\Delta_i$ are given in the main text. (b) Absolute values of the eigenfrequency difference $|\Omega_+ - \Omega_-|$ in the parameter space \((\theta_1, \Delta_1)\). Circles mark the positions of the EPs. (c) Argument of the eigenfrequency difference, used to compute vorticities as defined in Eq.~(\ref{eq:vorticity}). (d) Optical chirality of the eigenmodes [Eq.~(\ref{eq:chiral})]; red and blue indicate dominant counterclockwise (CCW) and clockwise (CW) angular momentum components, respectively. (e) Enlarged views of regions A, B, and C in (b), (c), and (d). Four selected loops, corresponding to Classes I–IV, are drawn with distinct curves. The range of the parameter is $(\theta_1,\Delta_1) \in [-0.127\pi, -0.072\pi] \times [0, 0.35R]$. The four EPs—EP$_a$, EP$_b$, EP$_c$, and EP$_d$—are located at $(\theta_1,\Delta_1) = (-0.115\pi, 0.126R), (-0.096\pi, 0.063R), (-0.084\pi, 0.137R)$, and $(-0.086\pi, 0.203R)$, respectively. (f) Spatial mode profiles of representative EPs labeled in (e)-C. Small arrows indicate the Poynting vector, computed as $\vec{s}(\mathbf{r}) = \text{Im}[\psi^*(\mathbf{r}) \vec{\nabla} \psi(\mathbf{r})]$, while the large curved arrows guide the direction of $\vec{s}$.}
    \label{fig:fig1}
\end{figure*}

\section{Realizations of topological structures in microcavity systems}
\label{sec:realization}

The physical realization of multi-EP topologies is demonstrated by examining a dielectric microdisk perturbed by three subwavelength scatterers. Fine-tuning the scatterer parameters in this platform enables the robust realization of multiple EPs with controllable topological features. Moreover, this setup allows for the design of parameter-space loops that selectively enclose EP pairs corresponding to the four distinct topological classes predicted by the two-level model. This system, thus, serves as a promising candidate for implementing and probing the interplay among vorticity, Berry phase, and eigenmode projection onto a basis state in a controlled and experimentally accessible manner \cite{Wiersig2014enhancing, Wiersig2016sensors, Peng2016chiral, Chen2017exceptional, Mao2024exceptional}.

\subsection{A microcavity with three scatterers}
\label{sec:systems}

To investigate the optical eigenmodes of a dielectric microdisk perturbed by three asymmetrically configured scatterers [see Fig.~\ref{fig:fig1}(a)], we solve the two-dimensional Maxwell equations, which, under the assumption of harmonic time dependence and TM polarization, reduce to a scalar Helmholtz equation:
\begin{equation}
    -\nabla^2\vec{\psi}(\mathbf{r}) = n^2(\mathbf{r})k^2\vec{\psi}(\mathbf{r}).
\end{equation}
Here, the electric field is oriented along the out-of-plane direction, $\vec{\psi} = (0, 0, E_z)$, and the continuity of both the field $\psi$ and its normal derivative $\partial_\nu \psi$ across disk boundaries is imposed~\cite{Jackson1999classical, Chang1996optical}. To solve for the eigenmodes numerically, we employ the generalized Lorenz--Mie theory~\cite{Gagnon2015lorenz, Gouesbet2011generalized}, which is particularly effective for handling circular disks that preserve rotational symmetry. At infinity, the pure-outgoing wave condition is assumed, so that the resulting eigenfrequencies are complex-valued: $\Omega \equiv kR = \omega R / c \in \mathbb{C}$, with the imaginary part of $kR$ quantifying the decay rate of modes. The refractive index is taken to be piecewise uniform: a constant $n > 1$ within the microdisk and $n_0 = 1$ outside. In the present study, we restrict ourselves to the case of $n = 2$, which captures the essential physics without loss of generality.

\begin{figure*}
    \centering
    \includegraphics[width=1.0\linewidth]{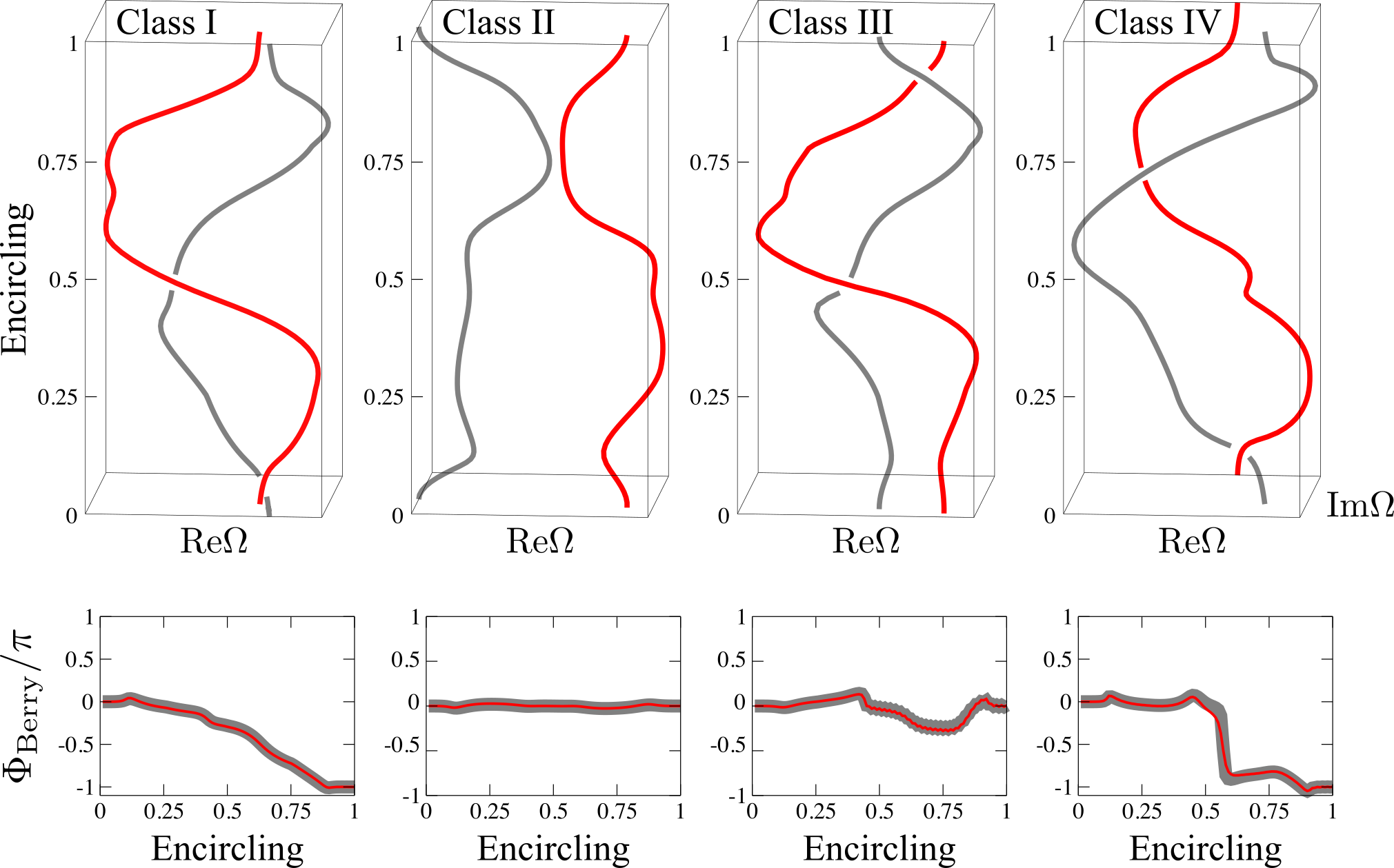}
    \caption{Braiding structures and accumulated Berry phases associated with four types of EP pairs, corresponding to Classes I–IV. The gray and red lines represent the two eigenmodes. (a) Class I: two EPs with opposite vorticities and opposite chiralities. The resulting braid degree is zero (trivial), and the accumulated Berry phases are \((\pi, \pi)\). (b) Class II: two EPs with opposite vorticities and the same chiralities. The braid degree remains zero, and the accumulated Berry phases are \((0, 0)\). (c) Class III: two EPs with the same vorticities and opposite chiralities. The braid degree is two (nontrivial), while the Berry phases remain \((0, 0)\). (d) Class IV: two EPs with the same vorticities and the same chiralities. The braid degree is two, and the accumulated Berry phases are \((\pi, \pi)\). These topological characteristics match those summarized in Table~\ref{tab01}.}
    \label{fig:fig2}
\end{figure*}

\subsection{Multiple EP pairs: vorticity and chirality}
\label{sec:MultipleEPpairs}

In the following, we focus on two basis eigenmodes that form a degenerate pair when the scatterers are not introduced. This degeneracy is intrinsic to the modes in a circular dielectric disk, as the modes are given by $\psi \sim J_m(nkr)e^{\pm i m \theta}$ [or equivalently, $J_m(nkr)\cos(m\theta)$ and $J_m(nkr)\sin(m\theta)$]. Namely, for an eigenfrequency obtained for a specific $m \ne 0$, the two corresponding eigenmodes form an orthogonal degenerate pair. However, when scatterers are introduced, these two eigenmodes couple and result in their superpositions. The complex-valued eigenfrequencies of the two coupled modes vary along a parameter-dependent Riemann surface.

We examine the eigenmode pair with $m = 16$ in a circular disk of radius $R$, perturbed by three scatterers whose fixed radii are $(r_1, r_2, r_3) = (0.04R, 0.05R, 0.06R)$, and whose positions are given by $\{(\theta_1,\Delta_1), (\theta_2,\Delta_2), (\theta_3,\Delta_3)\}$ [see Fig.~\ref{fig:fig1}(a)]. Here, we fix the positions of two scatterers as $\{(\theta_2,\Delta_2), (\theta_3,\Delta_3)\} = \{(\pi - \delta, 0.05R), (\pi + \delta, 0.05R)\}$, where $\delta = \pi / 20$. We then compute the eigenmodes by exploring the parameter space $(\theta_1, \Delta_1)$. As shown in Fig.~\ref{fig:fig1}, a total of 17 EPs are identified. The vanishing difference between the two complex-valued eigenfrequencies accurately marks the positions of the EPs [Fig.~\ref{fig:fig1}(b)]. To unveil their topological properties, we define four distinct loops that enclose various EP pairs [see Fig.~\ref{fig:fig1}(e)--I].

The vorticities, defined in Eq.~(\ref{eq:vorticity}), reveal the topological structures of the complex eigenfrequencies [Fig.~\ref{fig:fig1}(c)]. Faint white curves, linking two nearby circular symbols in Fig.~\ref{fig:fig1}(c), indicate real branch cuts, along which the Re($kR$) values coincide. Note that, in Fig.~\ref{fig:fig1}(e)-II, the pair EP$_b$ $-$ EP$_c$ are connected by a real branch cut, whereas the pairs EP$_a$ $-$ EP$_b$, as well as EP$_c$ $-$ EP$_d$, are connected by imaginary branch cuts [i.e., Im($kR$) values coincide]. The presence or absence of branch cuts between EP pairs reflects the connectivity of their spectral topology.

The chirality of the eigenmode propagation direction is defined as
\begin{align}
\label{eq:chiral}
    \text{Chirality} = \frac{\sum\limits_{m>0} |\alpha_m|^2 - \sum\limits_{m<0} |\alpha_m|^2}{\sum\limits_{m>0} |\alpha_m|^2 + \sum\limits_{m<0} |\alpha_m|^2} \ ,
\end{align}
where $\alpha_m$ denotes the expansion coefficient of the modes inside the disk: $\psi(r,\theta) = \sum_m \alpha_m J_m(n k r) e^{i m \theta}$. The parameter-dependent chirality is shown in Fig.~\ref{fig:fig1}(d). As shown in the figure, at the EPs the chiral optical modes exhibit highly asymmetric angular momentum components, i.e., the coefficients $\alpha_m$ are dominant either for $m > 0$ or for $m < 0$. When the positive (negative) components dominate, the corresponding mode exhibits counterclockwise (clockwise) rotation, denoted as CCW (CW). Our numerical results confirm that the EP modes possess nearly perfect chirality (greater than 0.99) in either the CCW (+) or CW (-) direction.

\subsection{Encircling EP pairs: braid and Berry phase}
\label{sec:EncirclingEPpairs}

When EP pairs are encircled along the selected loops, the complex eigenfrequencies exhibit braiding and the corresponding eigenstates accumulate Berry phases. The results are summarized in Table~\ref{tab01}.

We first consider EP pairs with opposite vorticities. In Class I [Fig.~\ref{fig:fig2}(a)], two EPs with opposite vorticities and opposite chiralities are encircled. The resulting braiding is trivial (braid degree = 0), while the accumulated Berry phases of the two spatial mode profiles are nontrivial, $(\pi, \pi)$. These phases arise from the total variation of the biorthogonal inner products of the eigenmodes during the adiabatic encircling of the EPs~\cite{Garrison1988complex, Mailybaev2005geometric, Cui2012geometric, Wagner2017numerical}. As the two EPs approach each other, they merge into a diabolic point.

In Class II [Fig.~\ref{fig:fig2}(b)], we encircle two EPs with opposite vorticities and the same chiralities. Although the vorticities remain the same as in Class I, the accumulated Berry phases are now trivial, $(0, 0)$, and the braiding remains trivial. When these EPs merge, their defectiveness persists, although the topological effects of braiding and Berry phase accumulation no longer manifest.

Next, we examine EP pairs with the same vorticities. In Class III [Fig.~\ref{fig:fig2}(c)], two EPs with the same vorticities and opposite chiralities are encircled. The resulting braiding is nontrivial (braid degree = 2), while the accumulated Berry phases remain trivial, $(0, 0)$. As the EPs merge, they form a vortex point, which is not defective.

Finally, in Class IV [Fig.~\ref{fig:fig2}(d)], we encircle two EPs with the same vorticities and the same chiralities. Although the vorticities are the same as in Class III, the accumulated Berry phases differ, yielding $(\pi, \pi)$. The braiding remains nontrivial (braid degree = 2), and the merged EPs retain their defectiveness.


\section{Summary}
\label{sec:summary}

We explore the topological structures arising from pairs of exceptional points in two-level non-Hermitian systems, where eigenvalues and eigenvectors simultaneously coalesce. We focus on two central topological features associated with such EPs: the braiding of complex eigenenergies and the accumulation of Berry phases during adiabatic encircling in parameter space.

To systematically characterize these phenomena, we classify EP pairs into four distinct classes according to three criteria: the vorticity and braids of the eigenenergies, the Berry phase accumulated upon encircling, and the eigenstate projection onto a basis state near each EP. For each class, we construct minimal two-level effective Hamiltonians with EP pairs and derive their associated braid structures and phase responses. This classification reveals how different configurations of EP pairs give rise to qualitatively distinct topological effects, even when the underlying energy spectra appear similar.

We then verify these classifications numerically in a physically relevant platform, an optical microdisk cavity with three scatterers. By solving the two-dimensional Maxwell equations for transverse magnetic modes using generalized Lorenz Mie theory, we identify multiple EPs in the system’s complex frequency spectrum. Through careful construction of loops in the parameter space that encircle various EP pairs, we observe the predicted combinations of braid degrees and Berry phase accumulations. The eigenmode projection onto a basis state near each EP is manifested in the form of chiral mode profiles with asymmetric angular momentum components, rotating either clockwise or counterclockwise. These chiral features further confirm the theoretical classification based on eigenstate projection onto a basis state.

Overall, we provide a unified, experimentally relevant framework for understanding the interplay of braiding, Berry phase, and eigenstate projection onto a basis state in non-Hermitian systems with multiple EPs. The results offer both conceptual insights into the classification of EP-induced topology and practical pathways toward their realization and control in photonic systems.

\acknowledgments
We acknowledge financial support from the Institute for Basic Science in the Republic of Korea through the project IBS-R024-D1.

\bibliography{reference}

\end{document}